\documentclass[pra,aps,showpacs,floatfix,preprint]{revtex4-1}
\usepackage{color}
\usepackage[utf8]{inputenc}

\usepackage{colordvi}

\usepackage{graphicx,amssymb,epsfig,amsmath}
\usepackage{epstopdf}

\usepackage{amsmath}
\usepackage{bm}
\usepackage{amsfonts}

\begin{document}

\title{Dynamics of Hubbard Hamiltonians with the multiconfigurational
  time-dependent Hartree method for indistinguishable particles}

\author{Axel U. J. Lode}
\email{axel.lode@unibas.ch} 
\author{Christoph Bruder}
\affiliation{Department of Physics, University of Basel, Klingelbergstrasse 82, CH-4056 Basel, Switzerland}

\begin{abstract}
We apply the multiconfigurational time-dependent Hartree method for
indistinguishable particles (MCTDH-X) to systems of bosons or fermions
in lattices described by Hubbard type Hamiltonians with long-range or short-range interparticle interactions. The wavefunction
is expanded in a variationally optimized time-dependent many-body
basis generated by a set of effective creation operators that are related to the original particle creation operators by a time-dependent unitary transform. We use the time-dependent variational principle for the coefficients of this transform as well as the expansion coefficients of the wavefunction in the time-dependent many-body basis as variational parameters to derive equations of motion. 
The convergence of MCTDH-X is shown by comparing its results to the exact diagonalization of one-, two-, and three-dimensional lattices filled with bosons with contact interactions.
We use MCTDH-X to study the buildup of correlations in the long-time splitting dynamics of a Bose-Einstein condensate loaded into a large two-dimensional lattice subject to a barrier that is ramped up in the center. We find that the system is split into two parts with emergent time-dependent correlations that depend on the ramping time -- for most barrier-raising-times the system becomes two-fold fragmented, but for some of the very fast ramps, the system shows revivals of coherence. 
\end{abstract}
\maketitle

\section{Introduction}
Ultracold atoms in optical lattices are a very active and wide field
of research that bridges the gap between atom-, molecular, and optical
physics and condensed matter systems. 
They have been employed as quantum simulators for condensed matter
systems that cannot be experimentally controlled to the same extent as
ultracold atoms in optical lattices \cite{CB98,BH,fer2}. Recent progress in this direction includes for instance the realization of artificial gauge fields \cite{gauge-exp} and topological states of matter \cite{top-exp} with cold atom systems. 

To describe the dynamics of ultracold atoms in optical lattices
theoretically requires solving the equation governing these systems,
i.e., the time-dependent many-body Schrödinger equation
(TDSE). However, 
the dimensionality of the many-body Hilbert space that hosts the solution grows exponentially with the number of particles \textit{and} with the number of sites in the considered optical lattice.
Since no general analytical solution is known to date, many approximate numerical methods have been devised. Popular approaches include the time-dependent density matrix renormalization group \cite{TDMRG}, matrix product states \cite{MPS1,MPS2,MPS3}, time-evolved block decimation \cite{TEBD1,TEBD2}, dynamical mean-field theory \cite{FDMFT,BDMFT,DMFT-1st,DMFT-MCTDH}, mean-field lattice methods, the discrete nonlinear Schrödinger equation \cite{DNLS}, and the Hartree-Fock theory for Hubbard Hamiltonians \cite{HubbardHF}.
These methods have problems when dealing with optical lattices of spatial dimension larger than one (time-dependent density matrix renormalization group, matrix product states, time-evolved block dimension), when the considered lattice is not spatially homogeneous (dynamical mean-field theory), or they oversimplify the emergent many-body correlations (discrete nonlinear Schrödinger equation and Hartree-Fock for Hubbard Hamiltonians).

For continuous systems of ultracold atoms, i.e.,  atoms that do
\textit{not} reside in an optical lattice, a theory which does not
suffer from the aforementioned problems has been
formulated~\cite{alon,MCTDHX} and
implemented~\cite{exact_F,MCTDH-X,S-MCTDHB}: the multiconfigurational
time-dependent Hartree for indistinguishable particles. 
In this work, we apply the same philosophy to describe ultracold atoms in optical lattices.

We adopt the strategy of Refs.~\cite{alon,MCTDHX} and apply the time-dependent variational principle \cite{DFVP,TDVP} to the time-dependent many-body Schrödinger equation using a formally complete, time-dependent, and variationally optimized orthonormal many-body basis set and demonstrate the exactness of the obtained theory, the multiconfigurational time-dependent Hartree method for indistinguishable particles in lattices (MCTDH-X). 
In the static case, i.e., the solution of the time-independent
Schrödinger equation, we compare MCTDH-X for a bosonic Hamiltonian
with exact diagonalization and find that the error in the obtained
method goes to zero roughly exponentially with the number of effective
time-dependent one-particle basis functions employed. In the dynamical
case, i.e., the solution of the time-dependent Schrödinger equation,
we apply MCTDH-X to the long-time  splitting dynamics of bosons in a
two-dimensional lattice. 
We show that correlations in the split system are built up almost independently of the splitting times: the reduced one-body density matrix acquires two eigenvalues on the order of the particle number throughout the splitting process, i.e., fragmentation emerges. Fragmentation emerges with a delay for short splitting times which becomes proportional to the splitting time for longer splitting times. Interestingly, revivals of the uncorrelated (coherent) initial state are seen for very short splitting times. By quantifying the coherence of the system with the first-order correlation function, we show that the mechanism behind the fragmentation is the loss of coherence between the left and right fractions of the bosons as the barrier is ramped up.

\section{Theory}
\subsection{Schrödinger equation and Hubbard Hamiltonians}
The aim is to solve the time-dependent many-body Schrödinger equation,
\begin{equation}
 i \partial_t \vert \Psi \rangle =  \hat{H} \vert \Psi\rangle,\label{TDSE}
\end{equation}
for the Hubbard Hamiltonian including a general long-ranged interaction,
\begin{equation}
 \hat{H}_{\text{Hub}} = \sum_j \epsilon_j \hat{b}^\dagger_j \hat{b}_j - J \sum_{<j,k>} \left[ \hat{b}^\dagger_j \hat{b}_k + \hat{b}^\dagger_k \hat{b}_j \right] + \frac{U}{2} \sum_{i\leq j} W(i,j) \left( \hat{b}^\dagger_i \right)^2\left( \hat{b}_j \right)^2. \label{HBH}
\end{equation}
Here and in the following, $<\cdot,\cdot>$ denotes neighboring sites. The full many-body state reads
\begin{equation}
 \vert \Psi \rangle = \sum_{\lbrace \vec{n} \rbrace} C_{\vec{n}}(t) \vert \vec{n} \rangle;\qquad \vert \vec{n} \rangle = \sqrt{\frac{1}{n_1! \cdots n_{M_s}!}} \left( \hat{b}^\dagger_1 \right)^{n_1} \cdots \left( \hat{b}^\dagger_{M_s} \right)^{n_{M_s}} \vert vac \rangle. \label{fullWF}
\end{equation}
The sum in this ansatz runs over all possible configurations of $N$ indistinguishable particles in the $M_s$ lattice sites. In the exact diagonalization approach, one takes Eq.~\eqref{fullWF} as the ansatz and determines the coefficients $C_{\vec{n}}(t)$.
However,
the number of the coefficients $C_{\vec{n}}(t)$ in the many-body ansatz Eq.~\eqref{fullWF} grows as a factorial [$\binom{N+M_s-1}{N}$ for bosons and $\binom{M_s}{N}$ for fermions], i.e. exponentially, with the number of lattice sites. 

To make large lattices tractable, approximations have to be introduced to reduce the number of coefficients. In what follows, we use a variational approach to obtain such an approximation: the general ``MCTDH ansatz'' is used to derive equations of motion. 

\subsection{MCTDH-X for Hubbard Hamiltonians}

The MCTDH-X equations of motion are derived with a time-dependent variational principle of the Schrödinger equation~\cite{TDVP} for Hubbard Hamiltonians by taking into account multiple ``effective'' time-dependent creation operators that act on all $M_s$ lattice sites.  

To start, we note that one can transform the time-independent operators $\hat{b}^\dagger_j$ in the Hubbard Hamiltonian in Eq.~\eqref{HBH} using a time-dependent unitary matrix $u_{kj}(t)$ to a set of $M$ operators $\lbrace a^\dagger_k(t) \rbrace$,
\begin{equation}
 \hat{a}^\dagger_k \equiv \sum_{j=1}^{M_s} u_{kj} (t) \hat{b}^\dagger_j,
\end{equation}
with the inverse transformation
\begin{equation}
 \hat{b}^\dagger_j = \sum_{k=1}^M u^*_{kj} \hat{a}^\dagger_k(t). \label{BH_DNLS_op}
\end{equation}
It is important to note that in the above equation $M$ is the number of \textit{effective creation operators} $\hat{a}_k(t)$ considered and \textit{is not the number of sites} $M_s$ in the system. 
To proceed, one forms all possible configurations $\vec{m}$ of $N$ particles in $M$ of the above effective time-dependent operators $\hat{a}^\dagger_k(t)$. 
The resulting ansatz for the many-body state then reads
\begin{eqnarray}
 \vert \Psi \rangle &=& \sum_{\vec{n}} C_{\vec{n}}(t) \vert \vec{n} \rangle \equiv \sum_{\vec{m}} C_{\vec{m}}(t) \vert \vec{m}; t \rangle\label{MC_BH_ansatz} \\
 \vert \vec{m};t \rangle &=& \sqrt{\frac{1}{m_1! \cdots m_M!}} \left( \hat{a}^\dagger_1(t) \right)^{m_1} \cdots \left( \hat{a}^\dagger_M(t) \right)^{m_M} \vert vac \rangle.
\end{eqnarray}
The number of coefficients $C_{\vec{m}}(t)$ that has to be accounted for is $\binom{N+M-1}{N}$ for bosons and $\binom{M}{N}$ for fermions. The number of coefficients is no longer directly dependent on the number of sites $M_s$ in the treated system.
Note that the multiconfigurational ansatz in Eq.~\eqref{MC_BH_ansatz} contains the mean-field type wavefunctions of the time-dependent Hartree-Fock for fermions and the so called discrete non-linear Schrödinger equation for bosons as special cases, for $M=N$ and $M=1$, respectively.
The action functional~\cite{TDVP,DFVP} for the time-dependent many-body Schrödinger equation~\eqref{TDSE} can now be (re-)formulated using the operator relation in Eq.~\eqref{BH_DNLS_op} and reads
\begin{equation}
 S[\lbrace u_{kj} \rbrace, \lbrace u^*_{kj} \rbrace ] = \int dt \left( \langle \Psi(t) \vert \hat{H}_{\text{Hub}} - i \partial_t \vert \Psi(t) \rangle - \sum_{pq} \mu_{pq}(t) \left[ \sum_j  u^*_{pj} (t) u_{qj}(t) - \delta_{pq} \right]  \right) .\label{Action}
\end{equation}
Here, the time-dependent Lagrange multipliers $\mu_{pq}(t)$ have been introduced to enforce the orthonormality of the column vectors of the transform $u_{pk}(t)$. This ensures that $u_{pk}(t)$ is indeed unitary for all times $t$.
To derive the equations of motion for $u_{pk}(t)$, we proceed by inserting the transformed operators given in Eq.~\eqref{BH_DNLS_op} in the above action, Eq.~\eqref{Action}:
\begin{eqnarray}
 & &\langle \Psi(t) \vert \hat{H}_{\text{Hub}} - i \partial_t \vert \Psi(t) \rangle= \\
 & &\langle \Psi(t) \vert \bigg[ \sum_j \epsilon_j \sum_{pq} u^*_{pj}(t) u_{qj} (t) \hat{a}^\dagger_p(t) \hat{a}_q(t) - J \sum_{<j,k>} \sum_{pq} u^*_{pj}(t) u_{qk}(t) \hat{a}^\dagger_p(t) \hat{a}_q(t) \nonumber \\
 &+& \frac{U}{2} \sum_{i\leq j} \sum_{prqs} W(i,j) u^*_{pj}(t)u^*_{rj}(t) u_{qj}(t) u_{sj}(t) \hat{a}^\dagger_p(t) \hat{a}^\dagger_r(t)\hat{a}_q(t) \hat{a}_s(t) \nonumber \\
&-& i \sum_j \sum_{pq}   u^*_{pj}(t) \partial_t u_{qj}(t) \hat{a}^\dagger_p(t) \hat{a}_q(t) \vert \Psi(t) \rangle - i \sum_{\vec{m}} C_{\vec{m}}^* \partial_t C_{\vec{m}} \bigg] \vert \Psi(t) \rangle. \label{EOM_Orbs_to_vary}
\end{eqnarray}
With the abbreviations,
\begin{eqnarray}
 T_{pq} &=& - J \sum_{<k,s>} u^*_{pk}(t) u_{qs}(t)  \label{Tels} \\
 V_{pq} &=& \sum_j \epsilon_j u^*_{pj}(t) u_{qj}(t) \label{Vels} \\
\vec{W}_{rs}  &=& U \sum_j u^*_{rj}(t) W(i,j)  u_{sj} (t) \label{locint} \\
W_{rs_{pq}}  &=& U \sum_{k,j}  u^*_{pk}(t) u^*_{rj}(t)W(j,k) u_{qk} (t) u_{sj} (t) \label{Wels} \\
\rho_{kq} &=& \langle \Psi(t) \vert \hat{a}^\dagger_p (t) \hat{a}_q(t) \vert \Psi(t) \rangle\:, \label{RDMels}
\end{eqnarray}
Eq.~\eqref{EOM_Orbs_to_vary} reads
\begin{eqnarray}
 \langle \Psi(t) \vert \hat{H}_{\text{Hub}} - i \partial_t \vert \Psi(t) \rangle &=& \sum_{pq} \rho_{pq} \left( T_{pq} + V_{pq} - i \partial_{t_{pq}} \right) \\ \label{EOM_Orbs_to_vary2}
 &+&\frac{1}{2} \sum_{pqrs} \rho_{prqs} W_{rs_{pq}} - i \sum_{\vec{m}} C^*_{\vec{m}}(t) \partial_t C_{\vec{m}}(t). \nonumber
\end{eqnarray}
In the following, we use the vector notation $\vec{u}_p(t)$ for the column vectors of $u_{kj}(t)$ and refer to these vectors as orbitals. Furthermore, we use bold math symbols to denote matrices of the quantities defined in Eqs.~\eqref{Tels} and \eqref{Vels}.
The variation of the action functional with respect to the orbitals $\lbrace \vec{u}^*_p(t) \rbrace $ can now be performed. Since the action takes on \textit{the same functional form} as in the case of MCTDH-X, compare Eq.~\eqref{EOM_Orbs_to_vary2} to Eq.~(10) in Ref.~\cite{MCTDHX}, the variation with respect to the orbitals $\lbrace \vec{u}^*_p(t) \rbrace $ also must have the same functional form as the equations of motion of MCTDH-X. They take on the form
\begin{eqnarray}
 i \partial_t \vec{u}_j (t) &=& \mathbf{\hat{P}} \left. \Bigg(  \left[\mathbf{T} + \mathbf{V} \right] \vec{u}_j(t) \right. \label{EOM_Orbs} \\ 
 &+& \left. \sum_{k,s,q,l=1}^M \lbrace \rho(t) \rbrace^{-1}_{jk} \cdot \rho_{ksql}(t) \vec{W}_{sl} \vec{u}_q(t) \right. \Bigg) ; \nonumber \\
 \mathbf{\hat{P}}&=& \mathbf{1} - \sum_{j'=1}^M \vec{u}_{j'} (t)  \left(\vec{u}^*_{j'} (t)\right)^T. \nonumber
\end{eqnarray}
for the orbitals and 
\begin{eqnarray}
 \mathcal{H}(t) \mathcal{C}(t) = i \partial_t \mathcal{C}(t); \qquad
 \mathcal{H}_{\vec{m}\vec{m}'}(t) = \langle \vec{m};t \vert \hat{H} - i \partial_t \vert \vec{m}' ; t \rangle \label{EOM_Coeffs}
\end{eqnarray}
for the coefficients, where a vector notation $\mathcal{C}(t)=\lbrace
C_{\vec{m}}(t) \rbrace$ was introduced. The equations for the
orbitals, Eq.~\eqref{EOM_Orbs}, are presented here using the
invariance property 
$\sum_j u_{qj}^* \partial_t u_{kj}=0$ that follows from the unitarity
of $u_{kj}$.

The usage of the variational principle to derive equations of motion for
multiconfigurational wavefunctions in lattices yields the same result
as in the continuum case,
but the equations contain a different and particular spatial representation of the potential and kinetic energy, namely the matrices $\mathbf{T}$ and $\mathbf{V}$ which encode that the considered many-body state lives in a lattice. The existing implementation of MCTDH-X \cite{exact_F,MCTDH-X,S-MCTDHB} is straightforwardly adapted to the numerical solution of the equations of motion, Eqs.~\eqref{EOM_Coeffs} and \eqref{EOM_Orbs}, the modifications concern $\mathbf{T}$ and $\mathbf{V}$ only. 

Note that the projector $\mathbf{\hat{P}}$ above will vanish as soon
as the basis $\{ \vec{u}_j; j=1,M\}$  is complete. This is
\textit{exactly} the case only, if the number of effective operators
in the multiconfigurational ansatz is taken to be the number of sites
in the lattice $M=M_s$. Then, the time derivative in the orbitals' equations of motion \eqref{EOM_Orbs} is zero and the case of exact diagonalization is recovered; Eq.~\eqref{EOM_Coeffs} becomes equivalent to solving the TDSE with the full many-body state \eqref{fullWF} and the Hubbard Hamiltonian \eqref{HBH}. 
Furthermore, the equations of motion of MCTDH-X for Hubbard Hamiltonians, \eqref{EOM_Orbs} and \eqref{EOM_Coeffs}, boil down to the standard mean-field methods, i.e. the discrete non-linear Schrödinger equation in the case of $M=1$ and bosons and the time-dependent Hartree-Fock equations of motion in the case of $M=N$ and fermions. MCTDH-X for Hubbard Hamiltonians can hence be seen as a systematic generalization of mean-field methods: if the number of effective operators $M$ approaches the number of sites $M_s$ in the system the Hilbert space spanned by them is the full possible Hilbert space of the lattice. The variational principle used in the derivation of the approximation guarantees that the error in the description is minimal at any given point in time and, furthermore, once convergence is achieved even for a number of orbitals $M<M_s$, the dynamics of the full many-body wavefunction is captured \cite{Axel_exact,Axel_book,exact_F}.

\subsection{Quantities of interest}
We introduce the quantities that will be used in the
remainder of our work to analyze the obtained results. For notational
convenience, we will write vectors as functions of position, for instance $\vec{u}_k \equiv u_k(\vec{r})$.
The reduced one-body density matrix reads
\begin{equation}
 \rho^{(1)}(\vec{r},\vec{r}';t) = \sum_{kq} \rho_{kq} u_k^*(\vec{r}';t) u_q(\vec{r};t). \label{RDM1}
\end{equation}
The diagonal of the reduced one-body density matrix is the one-body density,
\begin{equation}
 \rho(\vec{r}) = \rho^{(1)}(\vec{r},\vec{r};t) = \sum_{kq} \rho_{kq} u_k^*(\vec{r};t)  u_q(\vec{r};t). \label{rho}
\end{equation}

To investigate correlations between the atoms, we use the fragmentation and the normalized first-order Glauber correlation functions $g^{(1)}$ \cite{Glauber}.

Fragmentation is computed from the eigenvalues of the reduced one-body density matrix $\rho_k^{(NO)}$. These eigenvalues $\rho_k^{(NO)}$ are termed natural occupations and are obtained by diagonaling the reduced one-body density matrix, see Eq.~\eqref{RDMels}. In a system with multiple natural occupations on the order of the number of particles, fragmentation can by quantified by
\begin{equation}
 F(t)=\sum_{k=2}^M \rho_k^{(NO)}(t) = 1 - \rho_1^{(NO)}(t).
\end{equation}
The quantity $F(t)$ defines what fraction of the atoms is outside the single particle state that corresponds to the largest eigenvalue $\rho_1^{(NO)}$ of the reduced one-body density matrix.
The first-order Glauber correlation function,
\begin{equation}
 g^{(1)}(\vec{r},\vec{r}';t)= \frac{\rho^{(1)}(\vec{r},\vec{r}';t)}{\sqrt{\rho(\vec{r};t) \rho(\vec{r}';t)}}, \label{G1}
\end{equation}
is a measure
for the spatial coherence and shows how well a product state can
describe the reduced one-body density matrix
$\rho^{(1)}(\vec{r},\vec{r}';t)$ at $\vec{r},\vec{r}'$: if a product
state or mean-field description is applicable to the system, then
$\vert g^{(1)} \vert^2=1$ holds, and if a product state of mean-field
description is \textit{not applicable} then $\vert g^{(1)} \vert^2<1$
is true. Note that the reduced one-body density matrix
[Eq.~\eqref{RDM1}] can be transformed to momentum space by applying a
Fourier transform to the orbitals $u_k(\vec{r};t)$. From the reduced
density matrix in momentum space, the momentum density
$\rho(\vec{k};t)$ and momentum correlations
$g^{(1)}(\vec{k},\vec{k}';t)$ can be computed in analogy to Eqs.~\eqref{G1} and \eqref{rho}, respectively.

\section{Results}
We study the eigenstates and dynamics of bosons with contact interactions, i.e., $W(i,j)=\delta_{ij}$ in Eq.~\eqref{HBH}, in the following two Subsections.

\subsection{Comparison with exact diagonalization}\label{statics}
\begin{figure}
 \includegraphics[width=\textwidth]{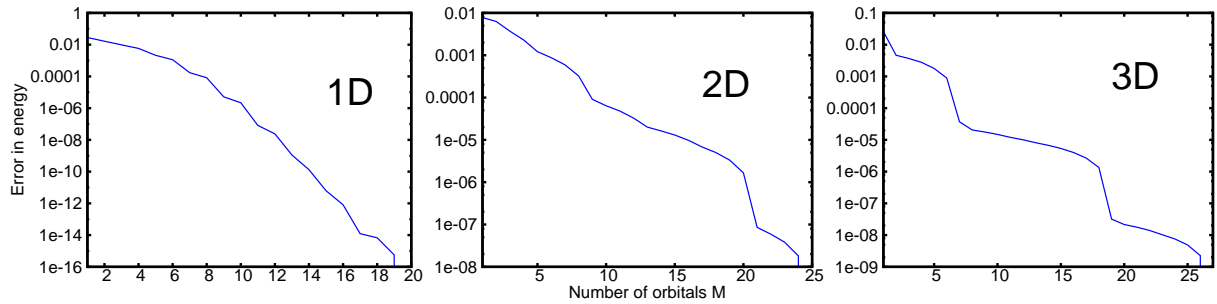}
 \caption{Comparison of MCTDH-X with exact diagonalization results. The relative error in the energies of $N=4,U=J=1$, harmonically trapped one-, two-, and three-dimensional bosons plotted in the left, middle, and right panel, respectively. In all cases the results converge roughly exponentially (note the logarithmic scale of the plots). All quantities shown are dimensionless.} 
 \label{Fig:energies}
\end{figure}

To demonstrate the correctness of our implementation of the MCTDH-X
method, we compare with
exact diagonalization results for a bosonic system in one-, two-,
and three-dimensional lattices of $20$, $5$ by $5$, and $3$ by $3$ by $3$ sites,
respectively. For the parameters in the Hamiltonian, Eq.~\eqref{HBH},
we choose a harmonic on-site energy $\epsilon_j= \frac{1}{2}
\omega^2 \vec{r}_j^2$ and an interparticle interaction of $U=J=1$. We
set the frequency $\omega$ of the external harmonic confinement to
$0.1$, $1$, and $3$ for the one-, two-, and three-dimensional
comparison with exact diagonalization,
respectively. Figure~\ref{Fig:energies} shows a plot of the error in
the energies as a function of the number of orbitals $M$ in the
computation. From the roughly exponential convergence of the error in
energy to zero, we infer that the derived method indeed recovers the
full complexity of the many-body state $\vert \Psi \rangle$ as the
number of variational parameters in the description is increased. This
means that the description of the Hubbard system with the introduced
effective creation operators $\lbrace \hat{a}^\dagger_k
\rbrace_{k=1}^M$ becomes complete and hence the solution of the
MCTDH-X equations of motion, Eqs.~\eqref{EOM_Orbs} \eqref{EOM_Coeffs},
is equivalent to the solution of the Schrödinger equation,
Eq.~\eqref{TDSE}. The time-dependent variational
principle \cite{DFVP,TDVP} and the MCTDH-X method
\cite{exact_F,Axel_exact,Axel_book,S-MCTDHB} imply the formal exactness of the approach.

\subsection{Dynamical splitting of a two-dimensional superfluid}\label{dynamics}
\begin{figure}
\includegraphics[width=\textwidth]{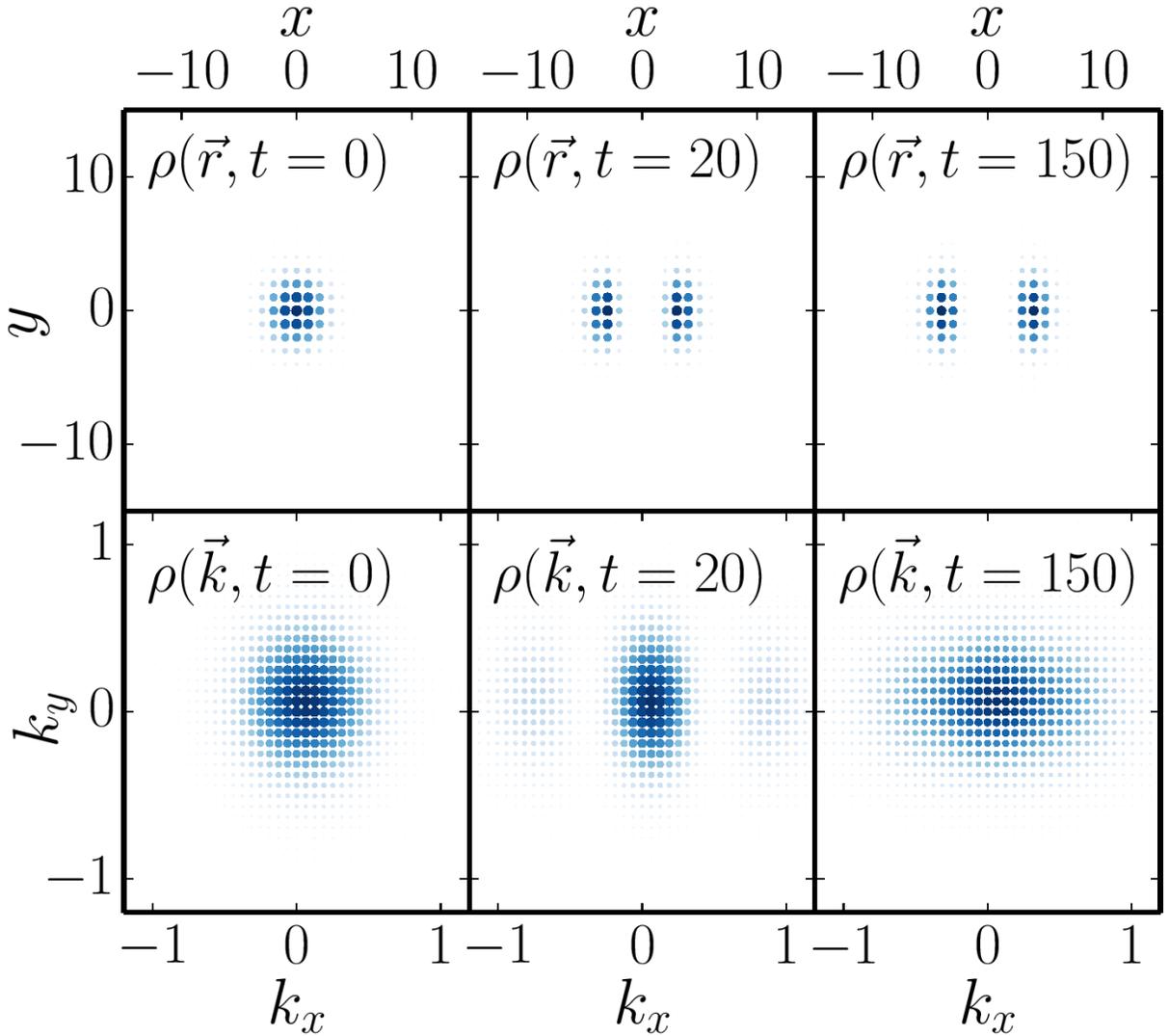}
\caption{Spatial and momentum densities in the splitting process with
  $t_{\text{ramp}}=50$. The top row depicts the spatial density
  $\rho(\vec{r};t)$ and the bottom row depicts the momentum density
  $\rho(\vec{k};t)$ for times $t=0,20$, and $150$,
  respectively. Darker color and larger points stand for larger
  on-site density. The plot of $\rho(\vec{r};t)$ shows that the
  initially Gaussian density distribution is split in two equal parts
  by the barrier. The left and right maxima of the density oscillate
  even for $t>t_{\text{ramp}}$ when the barrier has fully been ramped
  up: the process is non-adiabatic. The momentum density
  $\rho(\vec{k};t)$ exhibits an initially Gaussian
  distribution. During the splitting process, around $t=20$,
  $\rho(\vec{k};t)$ has three maxima which recombine to a broadened
  Gaussian distribution, once the splitting is complete and
  fragmentation has emerged [see $\rho(\vec{k};t=150)$ in the lower row and Fig.~\ref{Fig:frag}]. All quantities shown are dimensionless.}
\label{Fig:Densities}
\end{figure}

\begin{figure}
\includegraphics[width=\textwidth]{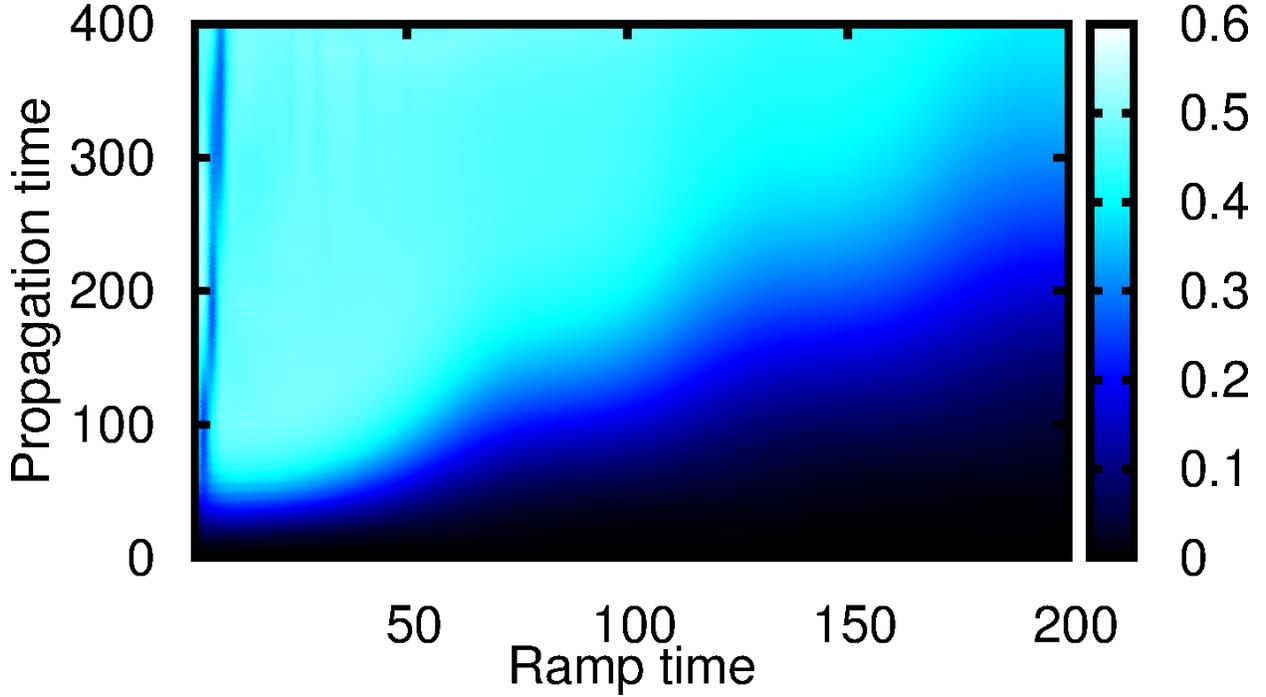}
\caption{Emergence of fragmentation in the splitting of a
  two-dimensional Bose-Einstein condensate in a lattice. The
  fragmentation of the system is shown as a function of the
  propagation time and the barrier-raising time
  $t_{\text{ramp}}$. Generally, fragmentation takes longer to set in
  for larger values of $t_{\text{ramp}}$, but there is a threshold for
  very fast ramps; for short barrier-raising-times
  $t_{\text{ramp}}\lesssim40$, the system needs about $50$ time units
  to become fragmented. For very short ramps $t_{\text{ramp}} \in
  \left[4,10\right]$, revivals of coherence are seen where the fragmentation returns back close to zero ($F=0$) after some propagation time. All quantities shown are dimensionless, see text for discussion.}
 \label{Fig:frag}
\end{figure}

\begin{figure}
\includegraphics[width=\textwidth]{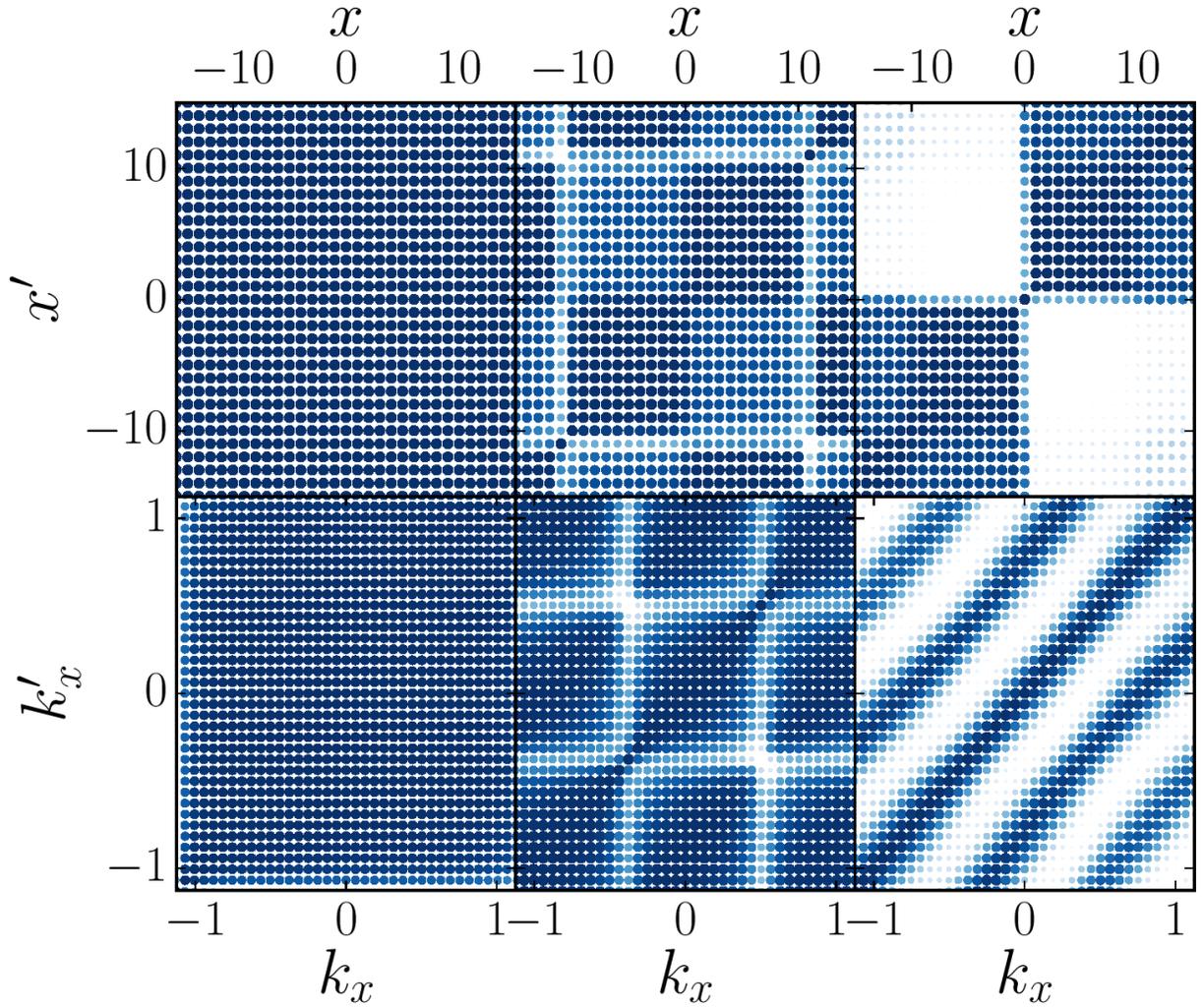}
\caption{Loss of spatial and momentum coherence in the splitting
  process with $t_{\text{ramp}}=50$. The top row shows the real-space
  correlation function $\vert
  g^{(1)}(\vec{r}=(x,0),\vec{r}'=(x',0);t)\vert^2$ and the bottom row
  depicts the momentum space correlation function $\vert g^{(1)}(\vec{k}=(k_x,0),\vec{k}'=(k_x',0);t)\vert^2$ for times $t=0,20$, and $150$, in the left, middle, and right columns, respectively. The darkest colors and biggest points correspond to $\vert g^{(1)}\vert=1$ (coherence) and white means $\vert g^{(1)}\vert=0$ (incoherence). Initially, at $t=0$, the bosons form a fully spatially- and momentum-coherent ensemble (see left column).
In the course of the splitting process, spatial coherence is maintained only within the formed minima of the double-well potential, while the coherence between the atoms in the left and right minima of the potential is lost and $\vert g^{(1)}\vert^2$ vanishes on the off-diagonal blocks (see emerging ``white squares'' in top middle and top right panels). In momentum space, the loss of coherence between the atoms in the left and right well shows by the emergence of a periodic diagonal stripe pattern (see lower row). All quantities shown are dimensionless.}
\label{Fig:Correlations}
\end{figure}

To assess that MCTDH-X can yield highly accurate predictions also for
dynamics in large lattices, we study the splitting dynamics of a
two-dimensional, initially parabolically trapped system of $N=100$
atoms in a $50$ by $50$ lattice with on-site repulsion $U=0.01J$. The size
of the configuration space of this system, $N_c = \binom{N+M_s-1}{N}=\binom{100+2500-1}{100}\approx4.7\cdot 10^{182}$ is far out of reach of exact diagonalization. In our simulations with MCTDH-X, we use $M=4$ effective operators and hence $\binom{N+M-1}{N}=\binom{100+4-1}{100}=176851$ configurations to describe the system. Since our results are converged with respect to the number of variational parameters, i.e., for the results for $M=3$ and $M=4$ all plots shown below are indistinguishable, our results can be considered as a numerically exact description of the on-going dynamics. The split is done by a Gaussian barrier in the center of the external harmonic confinement. The on-site energy offset is given by 
\begin{equation}
 \epsilon_j = \frac{1}{2} \omega_{\text{ext}}^2 \vec{r}^{2}_j + V(t) \exp{\left(-\frac{x_j^2}{2\sigma^2}\right)}.
\end{equation}
Here $\omega_{\text{ext}}$ is the frequency of the external harmonic trapping potential, $V(t)$ is the time-dependent height of the barrier and $\sigma$ its width.
For our simulations we choose the value $\omega_{\text{ext}}=0.3$, $\sigma=1$, and a linear ramp for the barrier $V(t)$,  
\begin{equation}
V(t) = \left\{\begin{array}{ll}  \frac{t}{t_{\text{ramp}}} V_{\text{max}} , & t\leq t_{\text{ramp}} \\
         V_{\text{max}}, & t > t_{\text{ramp}}
         \end{array}\right. .
\end{equation}
As a first step in our investigation, we choose an intermediate ramping time $t_{\text{ramp}}=50$, and analyze the time-evolution of the density. Fig.~\ref{Fig:Densities} shows a plot of the spatial and momentum densities at representative times. The spatial density $\rho(\vec{r};t)$ of the system is divided equally between the two wells by the split. Furthermore, the splitting is a non-adiabatic process, as there are remaining dynamics for $t>t_{\text{ramp}}$ after the split is complete: the maxima of the density in the left and the right minimum of the potential oscillate.
The (Gaussian) momentum density $\rho(\vec{k};t)$ is spread out in the $x$-direction. During the splitting process for $t\in [0,t_{\text{ramp}}]$, the momentum density $\rho(\vec{k};t)$ shows multiple peaks, similar to the peaks that are observed in the ground states of bosons in a double-well in continuous space at intermediate barrier heights (cf. Fig.~3 in Ref.~\cite{RDMs}).
We move on and monitor the emergence of fragmentation, i.e., the macroscopic occupation of more than one natural orbital in the splitting process as a function of the time within which the barrier is ramped up to its maximal value $V_{\text{max}}$ to split the Bose-Einstein condensate. 
To quantify fragmentation, we plot the fraction $F(t)$ of particles outside of the lowest natural orbital of the system for barrier-raising-times $t_{\text{ramp}}\in \left[2,200\right]$ in Fig.~\ref{Fig:frag}. 
Fragmentation emerges for all the investigated ramp times
$t_{\text{ramp}}$. The delay of the emerging fragmentation for longer ramp times is proportional to $t_{\text{ramp}}$, but constant at about $50$ time units in the case of
short ramps $t_{\text{ramp}}\lesssim40$. For very short ramps with
$t_{\text{ramp}}\approx 5$, revivals of the initial fragmentation
$F\approx0$ are seen in the dynamics. This bears some resemblance of
the \textit{inverse regime} discussed in Ref.~\cite{Split} for the
splitting of a one-dimensional Bose-Einstein condensate that does not reside in an
optical lattice potential, where it was found that the system stays
coherent counter-intuitively for larger interparticle repulsion at a
fixed $t_{\text{ramp}}$. In the present case of the splitting of a
Bose-Einstein condensate with an optical lattice potential, we find
that the system retains its coherence for very fast ramps at fixed and
small interparticle interactions. This behavior is 
counterintuitive as one would naively expect that a process that is further from 
being adiabatic and closer to a quench ($t_{\text{ramp}}\longrightarrow 0$) is more likely to break the coherence and drive the system to fragmentation. 
To get a space- and momentum-resolved picture of the emergence of
fragmentation and the entailed loss of coherence, we pick a ramping
time of $t_{\text{ramp}}=50$ and analyze the Glauber correlation
function $g^{(1)}$ in the course of the splitting process in
Fig.~\ref{Fig:Correlations}. Initially, at $t=0$ the atoms form a
single connected superfluid: all atoms are coherent and $\vert g^{(1)} \vert^2=1$ in both real and momentum space. During the splitting process, spatial coherence is preserved only within the left and within the right well, respectively. The spatial coherence between the atoms in the left and right well is lost as fragmentation emerges with time (see $\vert g^{(1)} \vert \approx 0$ regions in the top row of Fig.~\ref{Fig:Correlations}). Fragmentation shows also in the momentum correlation functions: throughout the splitting process, a periodic diagonal line pattern of alternating coherent and incoherent momenta emerges (bottom row of Fig.~\ref{Fig:Correlations}).

\section{Conclusions and outlook}
In this work, we formulated a general many-body theory to describe dynamics and correlations of indistinguishable interacting many-body systems in lattices. The software implementation of our theory is openly available as part of the MCTDH-X software package \cite{MCTDH-X}. 
We demonstrate that the method converges exponentially towards the exact result by comparing it to exact diagonalization. 

We obtain numerically exact results for the long-time splitting dynamics of initially condensed bosonic atoms in a large two-dimensional optical lattice: correlations that manifest as the macroscopic dynamical occupation of two natural orbitals are built up on a time-scale that is proportional to the barrier-raising-time $t_{\text{ramp}}$ -- the system becomes two-fold fragmented. Revivals of the initial coherence and absence of fragmentation is seen for very short ramping times. 

As an outlook, we mention a detailed investigation of the observed counter-intuitive revivals as well as the application of MCTDH-X to many-body systems of atoms with internal degrees of freedom \cite{S-MCTDHB} and/or systems subject to artificial gauge fields \cite{gauge-exp,top-exp} and spin-orbit interactions \cite{spielmanSOC,song:SOCfrag}.

\acknowledgments{Insightful discussions with Ofir E. Alon who supplied the initial idea for the theory presented in this manuscript are gratefully acknowledged. Comments and discussions about the manuscript with Elke Fasshauer and Marios C. Tsatsos, as well as financial support by the Swiss SNF and the NCCR Quantum Science and Technology and computation time on the Hornet and Hazel Hen clusters of the HLRS in Stuttgart are gratefully acknowledged.}

\clearpage

\end{document}